# Automatic Finite Elements Mesh Generation from Planar Contours of the Brain: an Image Driven, "Blobby" Approach.

Marek Bucki[1,2] & Yohan Payan[1,2]

[1] CMM Centro de Modelamiento Matemático, Av. Blanco Encalada 2120 Piso 7, Santiago de Chile, Chile
[2] TIMC Laboratory, UMR CNRS 5525, University J. Fourier, 38706 La Tronche, France

**Running title:**
Automatic FE Mesh Generation for Neurosurgery

**Corresponding author:**
Marek Bucki
CMM, Centro de Modelamiento Matemático
Av. Blanco Encalada 2120 Piso 7
Santiago de Chile
Chile

Tel: + 56 2 678 05 96
Email: mbucki@dim.uchile.cl

## Abstract

*In this paper, we address the problem of automatic mesh generation for finite elements modeling of anatomical organs for which a volumetric data set is available. In the first step a set of characteristic outlines of the organ is defined manually or automatically within the volume. The outlines define the "key frames" that will guide the procedure of surface reconstruction. Then, based on this information, and along with organ surface curvature information extracted from the volume data, a 3D scalar field is generated. This field allows a 3D reconstruction of the organ: as an iso-surface model, using a marching cubes algorithm; or as a 3D mesh, using a grid "immersion" technique, the field value being used as the outside/inside test. The final reconstruction respects the various topological changes that occur within the organ, such as holes and branching elements.*

## 1. Context

Accurate localization of the target is essential to reduce the morbidity during a brain tumor removal intervention. Image guided neurosurgery is nowadays established as an important development towards the future of neurosurgery. Nevertheless, it is still facing an important issue for large skull openings, with intra-operative changes that remain largely unsolved. In that case, deformations of the brain tissues occur in the course of surgery because of physical (dura opening, gravity, loss of cerebrospinal fluid, actions of the neurosurgeon, etc) and physiological phenomena (swelling due to osmotic drugs, anaesthetics, etc), some of them still being not completely known. Some studies have tried to measure this intra-operative "brain shift". Hastreiter and colleagues [1] observed a great variability of the brain shift ranging up to 24 mm for cortical displacement and exceeding 3 mm for the deep tumor margin; the authors claim for a non-correlation of the brain surface and the deeper structures. Nabavi and colleagues [2] state that the continuous dynamic brain shift process evolves differently in distinct brain regions, with a surface shift that occurs throughout surgery (and that the authors attribute to gravity) and with a subsurface shift that mainly occurs during resection (and that the authors attribute to the collapse of the resection cavity and to the intraparenchymal changes). As a consequence of this brain shift, the pre-operatively acquired images no longer correspond to reality; the pre-operative-based neuronavigation is therefore strongly compromised by intra-operative brain deformations.

In order to face this problem, authors have proposed to add to actual image-guided neurosurgical systems a module able to correct for brain deformations by updating the pre-operative images and planning according to the brain shape changes during surgery. The first algorithms developed for this proposed to deform the preoperatively acquired images using image-based models. Different non-rigid registration methods were therefore provided to match intra-operative images (mainly MRI exams) with pre-operative ones ([3-5]). More recently, biomechanical models of the brain tissues were proposed to constrain the image registration: the models are used to infer a volumetric deformation field from correspondences between contours [6-7] and/or surfaces [8] in the images to register. Arguing against the exorbitant cost of the intra-operative MRI imaging devices, some authors have proposed to couple the biomechanical model of the brain with *low-cost* readily available intra-operative data [9] such as laser-range scanner systems [10-11] or intra-operative ultrasound [12]. This proposal seems appealing from a very practical point of view, compared with the high cost intra-operative MRI device. However, it gives to the biomechanical model a crucial and very central position. This means that a strong modelling effort has to be carried out, concerning the design of the brain biomechanical model as well as its validation through clinical data.

This paper aims at introducing the methodology we propose to automatically build a patient-specific Finite Element (FE) model of the brain soft tissues gathering the aspects that are important from a modelling and from a clinical point of view, namely:
  (1) a full automatic procedure that generates patient-specific FE models with computational time constraints that are compatible with a clinical use,
  (2) models that can be generated from "sparse" imaging data such as low resolution MRI or non parallel ultrasound 2D slices,
  (3) FE models that can capture the topological changes inside the brain, such as the presence of ventricles,
  (4) FE models that can take into account the variability in the mechanical properties of the brain tissues,
  (5) finally, FE models that are mainly made of hexahedral elements known as being more accurate than tetrahedral elements.

After resuming the methods used in the literature for building biomechanical models of the brain, in the context of brain-shift neuronavigation, our methodology is described and discussed.

## 2. Available FE models of the brain, in the CAS context

Whereas the internal sub-structures of the brain (and the corresponding topological changes) probably have a strong influence onto the global mechanical behavior of the brain (for example, the interactions between the ventricles, the cerebrospinal fluid and the parenchyma) the biomechanical models first proposed in the literature only represented the brain tissues as an uniform isotropic structure. Finite Element models (with a 3D tetrahedral mesh) were therefore developed, assuming a linear small deformation framework with a single Young modulus value [13-14]. Miga and colleagues proposed a biphasic FE model where the brain is modeled as an elastic body (3D tetrahedral mesh) with an interstitial fluid [15]. Kyriacou and colleagues [6] were the first to propose a FE model that takes into account the topological changes due to internal sub-structures (the ventricles) as well as the internal rheological variability (with different Young modulus between white and grey matters). Unfortunately, the authors limited their work to a 2D model, arguing for the difficulties in the 3D mesh generation process needed to take into account the topological changes. Ferrant and colleagues proposed a FE model of the brain that is the most elaborate from the mesh topological point of view [16]. Indeed, starting from segmented 3D MR images, a patient-specific tetrahedral FE model is built, integrating key boundary surfaces, namely the cortical and the lateral ventricles surfaces. The algorithm proposed by Ferrant et al. can be seen as the volumetric counterpart of a marching tetrahedral iso-surface generation algorithm. The main limits of this algorithm are (1) that it needs regular and parallel labeled 3D images, (2) that it doesn't take into account the internal rheological variability and (3) that the mesh is made of tetrahedral elements. Indeed, as we mentioned above, it could be interesting to propose a FE model generation algorithm that takes as input "rough" imaging data such as low resolution MRI or non parallel ultrasound 2D slices. Moreover, hexahedral elements are known as being more accurate than tetrahedral elements in the context of FE analysis [17].

# 3. Our method

We propose to keep the idea of the FE mesh generation based upon an iso-surface introduced by Ferrant. We first define a scalar field generated by a set of outlines of the organ segmented from the volume patient data. This scalar field has an analytical expression and the organ surface is an iso-surface within it. The topological features of the brain, such as hemispheres independency and ventricles, are preserved. The FE mesh is generated using a grid-based technique: an initial regular grid is adapted to the organ surface defined by the iso-surface.

## 3.1 Definition of the scalar field

Our method defines a scalar field generated by outlines in 3D (or segments in 2D) based mainly on angle measure. Combining the fields generated by multiple outlines (or segments) we can obtain surfaces (or curves) that respect the following constraints:
- In 2D: all curves pass by the extremities of the segments.
- In 3D: all surfaces pass by the outlines.
- All curves and surfaces have a global tangent continuity.
- The orientation of the curves or surfaces can be constrained at the vicinity of the segment extremities, in 2D, or along the outlines, in 3D.

These properties along with the blending capabilities of the scalar field allow reconstructing the organ surface with respect to its topology. The complex structure of the organ is modeled by significant contours segmented within the volume data-set, and arranged in parallel slices. The branching problem raised by the changing topology between two consecutive slices is solved using surface orientation information extracted from the volume data. Compared to a marching cubes approach [18], the outlines based reconstruction allows focusing on the relevant parts of the organ, defined by the contours, while marching cubes generate surface patches around unwanted or not organic structures.

In the next section, the 2D formulation of the method is provided. The transition from 2D to 3D is quite straightforward and merely requires a natural extension to the third dimension of the very basic concepts used in 2D. Section 3.1.2 gives the detail of the 3D formulation. Volumetric mesh generation is described in section 3.2. We conclude and give an overview of future developments in section 4.

### *3.1.1 2D formulation*
*a) Definition*
Let's consider a linear segment [A, B]. We can define a 2D scalar field F as follows:

(1)     Let P be a point of the plane, F (P) = (angle between PA and PB) / $\pi$

In this definition the angle is always taken positive, in the range [0, $\pi$].

This field is well defined and continuous for any point of the plane, excepted in A and B. On any point on the opened segment ]A, B[ its value is 1. On any point of the infinite line (A, B) outside the segment [A, B] its value is 0. Outside the line (A, B) its value varies continuously within the range ]0, 1[.

A "λ-iso-line" (or "λ-iso-surface") within the field F is defined as a set of points P of the plane (or space) such as F(P)= λ. λ is called the "threshold" of the iso-curve. The field F is depicted in the Figure 1 along with 3 iso-lines corresponding to the thresholds ¾, ½ and ¼.

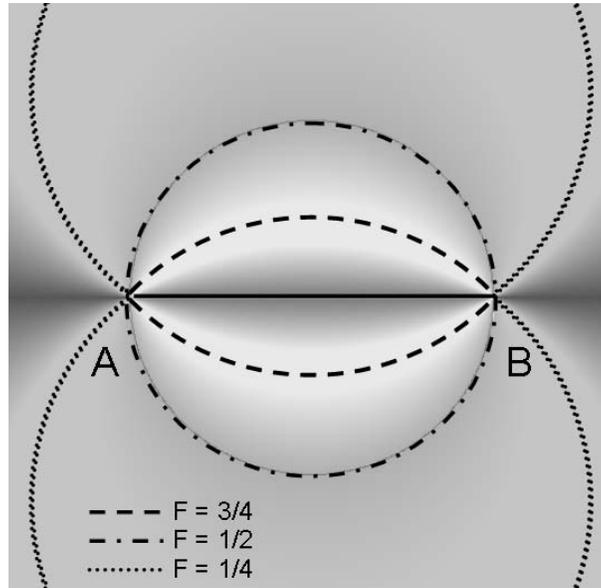

**Fig. 1:** scalar 2D field and three specific iso-lines

The discontinuity of the field in A and B allows extending continuously each iso-line by adding the points A and B. Yet these extended iso-lines present a tangent discontinuity in A and B.

This tangent continuity problem is solved by stitching 2 iso-lines with different threshold values. In the case of Figure 1, all λ–iso-lines on the side above the axis (A, B) match (1-λ)-iso-lines on the side below. Their union is a circle passing by A and B. In the rest of the text, "λ-iso-line" makes reference to such an extended and composite iso-line, with a λ threshold on the upper side of the segment and 1-λ on the lower side.

The relation between the tangent to the λ-iso-line in A and the value of λ is depicted in Figure 2. At the vicinity of A, point B is seen as infinitely distant. The angles defining the iso-line orientation are measured between each tangent (pointing towards each side of the segment [A, B]) and a horizontal vector pointing outwards the segment.

Let α be the angle formed in A between the upper tangent to the λ-iso-line and the horizontal vector. We have the following relation:

(2)  $\alpha = \lambda * \pi$

On the lower side of the segment [A, B], for the same tangent orientation, the measured angle is π–α, corresponding to a threshold value of 1-λ.

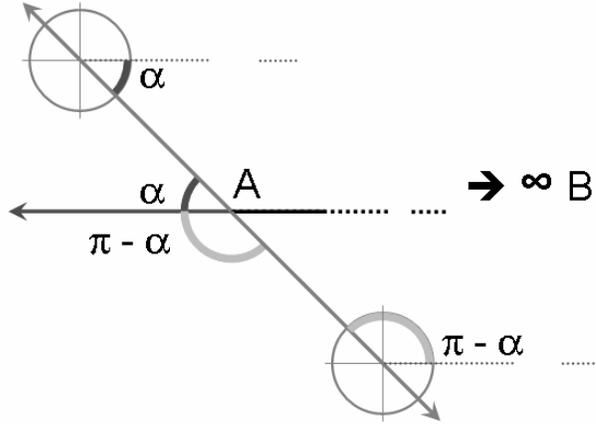 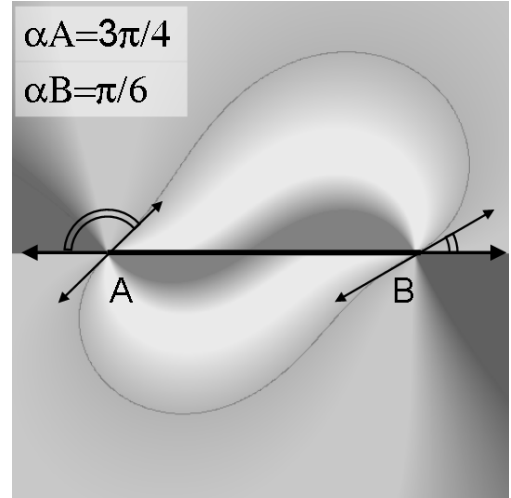

**Fig. 2:** angle measures at the vicinity of A     **Fig. 3:** scalar field with orientation control in A and B

It is easy to produce a scalar field, based on the one defined above, which allows us to control the orientation of a given iso-line at the vicinity of A and B. To achieve this we use "threshold varying" iso-lines, replacing the previous definition { P / F(P) = λ } by the following one { P / F(P)-λ(P) = 0 }, which is equivalent to considering 0-iso-lines (λ=0) within the field F-λ.

Let αA be the tangent angle in A and αB be the tangent angle in B. These two angles have corresponding threshold values λA=αA/π and λB=αB/π. We assume that A ≠ B. For any point P of the plane, let d(P,A) be the distance between P and A, and let d(P,B) be distance between P and B. We define the new field as follows:

(3)     $F_{λA,λB}(P) = F(P) - λ(P)$

Where F is the initial field as defined in (1) and λ is a threshold interpolation function defined by:

(4)     If point P is on the upper side of the segment:
        λ(P) = (λA*d(P,B) + λB*d(P,A)) / (d(P,A)+d(P,B))
        If point P is on the lower side of the segment:
        λ(P) = ((1-λA)*d(P,B) + (1-λB)*d(P,A)) / (d(P,A)+d(P,B))

We have λ(A)=λA, λ(B)=λB and between the two points the value is continuously interpolated.
The 0-iso-line within the field $F_{λA,λB}$ has the following properties:
- The curve passes through both points A and B.
- The tangents to the curve in points A and B form with the outgoing vector angles respectively αA and αB, measured on the upper side of the segment.

Figure 3 shows an example of the new scalar field computed with tangent orientation angles (αA, αB) being (3π/4, π/6), as well as the resulting 0-iso-line.

*b) Combining fields generated by parallel segments*

We want to generate a 0-iso-line that links together a set of segments, organized in parallel levels, respecting the orientation constraints given at the segments tips. Each level is a line, say horizontal, and the segments of a given level do not intersect. For a given point on the plane, the resulting field is obtained by "blending" the field generated by the segments of the level immediately above and those of the level immediately below the point.

For i=1..n, let $S_i=[E_{2i-1}, E_{2i}]$ be the n segments on a given level. These n segments define a set of 2n extremities $\{E_i\}_{i=1..2n}$. For i=1..n, let $(\lambda_{2i-1}, \lambda_{2i})$ be the thresholds given by (2) for each tangent angle defining the local surface orientation at $E_{2i-1}$ and $E_{2i}$. We have, thus, a set of 2n thresholds $\{\lambda_i\}_{i=1..2n}$. Let P be a point in the plane. The field generated by a given level is:

(5) $\quad F_{Level}(P) = (\Sigma_{i=1..n} F_i(P)) - \lambda(P)$

Where $F_i$ is the initial field associated to segment $S_i$, as defined in (1), and $\lambda$ is a threshold interpolation function which can be written as:

(6) $\quad$ If point P is on the upper side of the level:
$\quad\quad\quad \lambda(P) = \Sigma_{i=1..2n} (\lambda_i * \Pi_{j\neq i} d(P,E_j)) / (\Sigma_{i=1..2n} \Pi_{j\neq i} d(P, E_j))$
$\quad$ If point P is on the lower side of the level:
$\quad\quad\quad \lambda(P) = \Sigma_{i=1..2n} ((1-\lambda_i) * \Pi_{j\neq i} d(P,E_j)) / (\Sigma_{i=1..2n} \Pi_{j\neq i} d(P, E_j))$

Once the field computed for each level surrounding a given point P, the interpolation can be done in the following way. Let d(P,Up) be the distance between the point and the line supporting the segments of the upper level. Let d(P,Low) be the distance between the point P and the lower level. Let d(Up,Low) be the distance between the two levels. Let $F_{Low}$ and $F_{Up}$ be the field generated by each level. Let H be a continuous interpolation function defined on the interval [0, 1] which values range in [0, 1], such as H(0) = 1, H(1) = 0 and H(t) + H(1-t) = 1, for all t in [0, 1]. The total field value is:

(7) $\quad F_{Total}(P) = H(d(P,Up)/d(Up,Low))*F_{Up}(P) + H(d(P,Low)/d(Up,Low))*F_{Low}(P)$

Figure 4 shows an interpolated field for 3 segments (2 above and 1 below). In (4a) the field is generated with all tangents constrained to vertical. In (4b) the tangents are constrained to a 45° orientation towards the upper right corner. We can see the impact of this local orientation on the branching solution found by the algorithm. In (4a) the ambiguous branching situation leads to an "all-connected" solution whereas in (4b) it is more likely to find a diagonal connection between the upper-right and the lower segment.

The scalar field, as formulated above, has two drawbacks which we call "field shrinking" and "unwanted blending". The first is due to the diminution of the angular value in points remote from the segment. The effect is incorrect surface curvature. The latter effect generates incorrect connections between the outlines when two segments of the same level are too close. In 2D both effects have been suppressed. A "field expansion" function enhances the field strength in the proper connection direction and it weakens it in other directions, suppressing field shrinking (Figure 5a) as well as unwanted blending (Figure 5b).

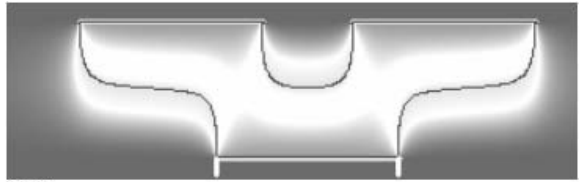
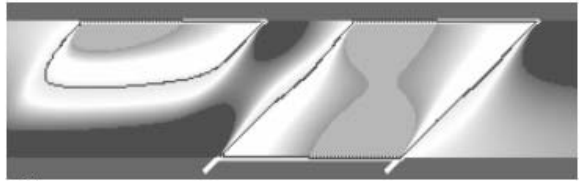
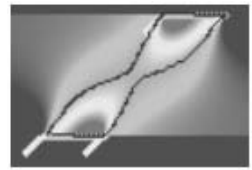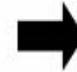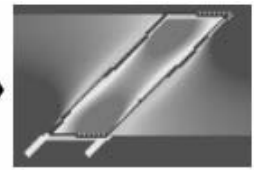
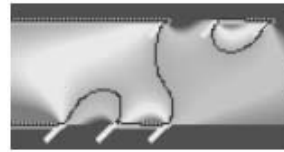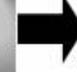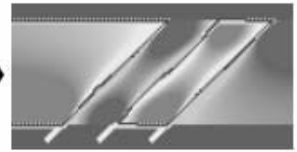

**Fig. 4:** field orientation and branching          **Fig. 5:** eliminating field shrinking and unwanted blending

*3.1.2  3D formulation*
The extension of the method to the third dimension is quite straightforward:
- **Segment** becomes **outline** i.e. a planar, closed, finite polygonal or parametric curve that does not intersect itself.
- **Segments on parallel levels** become **outlines on parallel slices.**
- **Vector pointing away from the segment** (as seen in Fig. 2 and 3) becomes **normal to the outline**, pointing towards the outside of the area defined by the closed curve, within the plane defined by the outline.
- **Angle** becomes **solid angle**. The solid angle of a region seen from a point P is the area of the unit sphere surface intersected by the cone whose summit is P and whose base is the outline of the region.
- **Curve** becomes **surface.**

Let's consider a planar region R enclosed by an outline. We can define the 3D scalar field F as follows:

(8)    Let P be a point in space, F (P) = (solid angle of region R seen from point P) / $2\pi$

We can see many similarities with the 2D field. $2\pi$ is half of the total unit sphere surface, just as $\pi$ was half of the total circle perimeter. The field is well defined and continuous for any point in space except for points lying on the outline of the region.
Points lying on the region plane, inside R, define a flat opened cone which intersects half of the unit sphere. The field value is then 1. Points lying on the region plane, outside R, define a flat closed cone whose solid angle measure is 0. The field value is 0 as well.
In the same way as for the 2D curves, the field discontinuity allows all the $\lambda$-iso-surfaces to be continuously extended by including the outline.

The λ parameter defines the tangent angle in the same way as in 2D and the tangent continuity problem is solved similarly by stitching together, on both sides of the area plane, 2 iso-surfaces with complementary thresholds λ and (1-λ).

The surface tangent orientation is depicted in Figure 6. In point P, the outline normal vector N1 and the area plane normal N2 define a local "outline normal plane" within which the surface tangent angle α is measured.

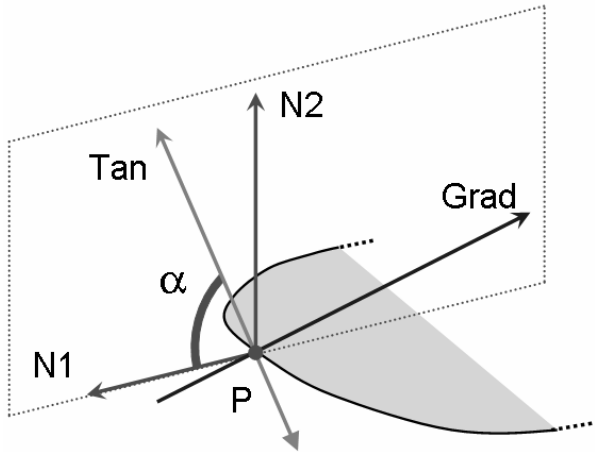
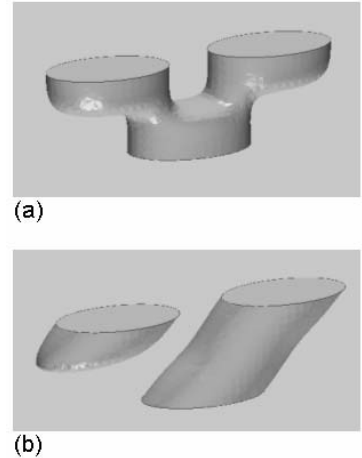

**Fig. 6:** local surface tangent orientation         **Fig. 7:** surface reconstruction from parallel outlines

In the context of neurosurgery, a pre-operative CT and/or MRI patient exam is available. From this exam parallel slices are extracted then organ segmentation yields contours within each slice. The organ surface local orientation can also be computed from the volume data. First the volumetric image gradient at point P is computed, yielding the vector Grad. We consider that the local organ tangent plane is the one containing point P and having Grad as normal vector. The intersection line of this plane with the outline local normal plane is computed. The tangent vector is a unitary vector supported by this line. Once we have computed the angle α, we can associate at point P a λ value that defines the local surface orientation.

In practice, the outlines yielded by manual or automatic segmentation are not parametric curves but rather closed polygonal lines. They do not have a continuous derivate. At each polygon node, the segment before and the segment after form an angle that is not 180°. A correction has to be applied to the λ value in order to take into account the local outline inflexion which has an impact on the solid angle measured at the vicinity of the node. The corrected threshold is the solid angle at which the "polygonal corner" is seen when converging towards the polygonal node at the given angle α, divided by 2π.

Similarly as in 2D, fields generated by parallel outlines can be combined using the equations (5) and (6) after only substituting the 2D field defined in (1) by the 3D one defined in (8). Furthermore, for any point between two consecutive and parallel slices, the resulting field is an interpolation of the fields generated by the immediate lower and upper outlines set. Equation (7) can be applied without any change.

Figure 7 shows an example of surface reconstruction from multiple outlines in a situation similar to the one illustrated in Figure 4. Two surfaces are generated by the same three elliptical outlines, two on the upper level and one below. In (7a), all surface tangents are constrained to vertical, which creates an ambiguous branching situation resolved by a global connection between the 2 levels. In (7b) the tangents are those of the cylindrical surfaces passing through the outlines, oriented 45º towards the upper right corner. This orientation information is used to solve the branching problem: the right hand side cylinder is fully reconstructed between the levels while the left hand side outline yields a partial tubular structure.

**3.2 Volumetric mesh generation**

In our application, the volume mesh is built using the 3D field described above. The brain outlines are placed within a regular hexahedral elements grid. At each grid node, the field value gives inside/outside information. Elements whose nodes are all inside are kept hexahedral. Elements which intersect the brain surface are subdivided in order to match the surface curvature. The intersection points between the reconstructed brain surface and the grid are computed using dichotomy. Each hexahedral element is divided into 6 pyramids, the pyramids summit being the center of the element, and the bases, each 6 sides of the hexahedron. Each pyramid is further subdivided into other elements, depending on its position respectively to the organ boundary.

The generated mesh takes into account the various topological changes that occur within the volume of the segmented brain allowing us modeling independence between both hemispheres, as well as cavities filled with cerebrospinal fluid.

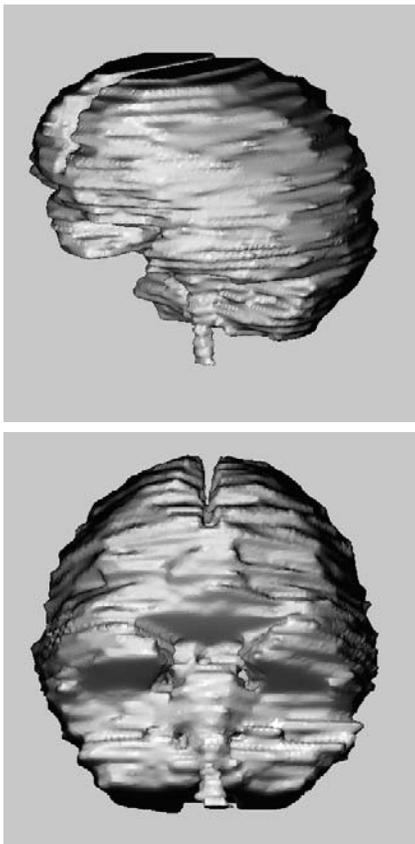

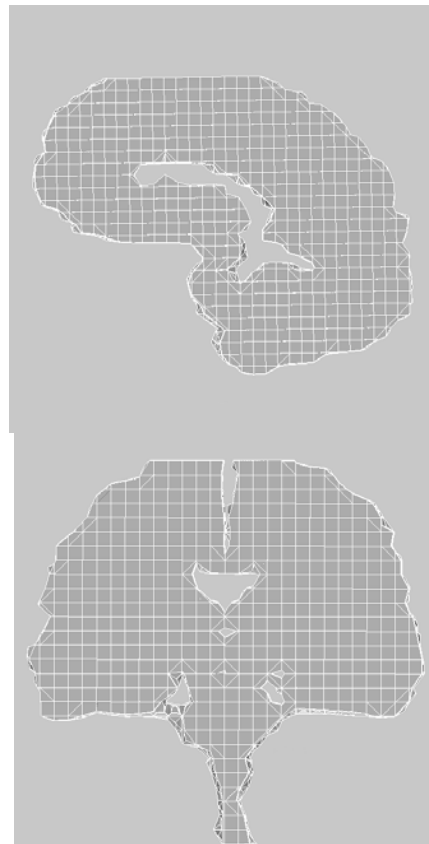

**Fig. 8:** mesh reconstructed from manually and partially segmented patient data

**Fig. 9:** sagittal and coronal slices within the FEM mesh shown on Fig. 8

# 4. Conclusion

The method was implemented and tested on patient data. It proved to be efficient enough to achieve the initial objective: an automatic FE mesh generation of the brain, based on segmented slices, which takes into account the topological changes within the organ.

With the introduction of automated MRI segmentation algorithms, such as those proposed in [19], in the data processing pipeline, we hope to develop a system allowing fully automated per-patient mesh generation.

Some aspects of the 3D method are being worked upon. While the "field shrinking" and "unwanted blending" issues have been solved in 2D, a new formulation of the solid angle is necessary in order to include a "field expansion" function. The impossibility of transposing directly the 2D method to 3D can be explained by the more complex outlines convexity conditions that make the field expansion direction more delicate to compute.

Our image based reconstruction needs a reliable object surface orientation computation algorithm. The actual implementation merely computes the grayscale gradient within the image volume. This is sufficient in the peripheral regions where the brain tissues can be easily delimited but the results are very noisy in the areas where this separation is unclear, such as the contact region between the two hemispheres.

The extension of the method to non-parallel slices is also being considered in order to apply it to 2.5D volume data yielded by a 2D echographic sweep of an organ (see [12]).

Finally a smarter mesh generation algorithm needs to be implemented in order to reduce the number (while increasing the size) of the inner elements of the mesh. The ideal would be to use an iso-surface based, hexahedral adaptative mesh generation algorithm such as the one proposed by Zhang et al. in [20].